\documentclass[12pt]{article}
\usepackage[pdftex]{graphicx} 
\usepackage{amsmath}
\usepackage{amssymb}
\usepackage{color}
\usepackage{simplewick}
\usepackage{slashed}
\usepackage{ulem}
\usepackage{mathrsfs}
\usepackage{ascmac}
\usepackage{amsthm}
\usepackage{physics}
\renewcommand{\theequation}{\arabic{section}.\arabic{equation}}

\counterwithin{equation}{section}

\begin{document}
\begin{titlepage}
\begin{flushright}
TIT/HEP-709\\
April, 2026
\end{flushright}
\vspace{0.5cm}
\begin{center}
{\Large \bf Integrals of motion in $WE_6$ CFT and the ODE/IM correspondence}

\lineskip .75em
\vskip 2.5cm
Daichi Ide, Katsushi Ito and Wataru Kono

\vskip 2.5em
 {\normalsize\it 
Department of Physics, Institute of Science Tokyo,
Tokyo, 152-8551, Japan\\

}
\vskip 3.0em
\end{center}
\begin{abstract}
We study the ODE/IM correspondence for the ordinary differential equation associated with the affine Lie algebra $E_6^{(1)}$. The WKB expansion of the solution of the ODE is performed by the diagonalization method, and the period integrals of the WKB coefficients along the Pochhammer contour are calculated. We also compute the integrals of motion on a cylinder in two-dimensional conformal field theory with W-symmetry associated with $E_6^{(1)}$. Their eigenvalues on the highest-weight state are shown to agree with the period integrals up to the sixth order.
\end{abstract}
\end{titlepage}


\section{Introduction}
Integrable field theories have attracted considerable attention, as they provide non-trivial examples of exactly solvable models. Integrability implies the existence of infinitely many conserved quantities, which are also referred to as the integrals of motion. In two-dimensional quantum field theories, the higher spin conserved charges constrain the dynamics of the theory, leading to the factorization of the scattering process into two-body scatterings, which are determined by the Yang-Baxter relations \cite{ZAMOLODCHIKOV1979253}. Integrals of motion, which are characterized by a hierarchy of soliton equations \cite{Drinfeld:1984qv,OLIVE1983491}, are also studied in the context of classical field theory. 
The affine Toda field theories \cite{Mikhailov:1980my,Hollowood:1989cg} are notable examples of integrable field theories with massive particles.

In two-dimensional conformal field theories (CFT), an infinite number of integrals of motion are found in \cite{Sasaki:1987mm,Eguchi:1989hs,Kupershmidt:1989bf}. In particular, the remarkable integrable structure was found in \cite{Bazhanov:1994ft,Bazhanov:1996dr,Bazhanov:1998wj}, where the family of mutually commuting operators is constructed from the monodromy operators in the quantum version of the KdV hierarchy.

The ODE/IM correspondence provides an interesting connection between ordinary differential equations (ODEs) and quantum integrable models \cite{Dorey:1998pt,Bazhanov:1998wj}. In this correspondence, the spectral problem of certain Schr\"odinger equations relates to the integrable structure of CFT via the functional relations. In particular, the Stokes coefficients of the connection problem of differential equations correspond to the transfer matrices and the Q-operators of the integrable model.
For the Schr\"odinger equation with monomial potential, certain WKB periods coincide with the integrals of motion of the Virasoro minimal models.
The correspondence for general polynomial-type potential can be further confirmed by the relation between the Thermodynamic Bethe ansatz equation and the exact WKB periods \cite{Ito:2018eon}.

The ODE/IM correspondence has been generalized to the relation between higher-order ODEs and CFTs with higher-spin fields. In particular, the correspondence between the linear differential equation associated with an affine Lie algebra and the CFT with W-algebra symmetry was studied \cite{Dorey:2006an,Ito:2015nla,Masoero:2015lga}. The relation for the affine Lie algebra $\hat{\mathfrak g}$ has been studied by using the Non-linear Integral Equation (NLIE) satisfied by the Q-functions \cite{Dorey:2006an,Ito:2020htm}.
The ODE is characterized by the order of the monomial potential and the generalized angular momenta. The results imply the correspondence to the CFT with W-algebra associated with the Langlands dual $\hat{\mathfrak g}^{\vee}$, where the order of the potential labels the non-unitary minimal series and the angular momenta are proportional to the momenta of the primary field in the free field representation.

The ODE/IM correspondence can be confirmed exactly by investigating the relation between the integrals of motion in CFT and the WKB expansion of the Stokes coefficients. So far, the $W_3$-CFT \cite{Zamolodchikov:1985wn,Fateev:1987vh} has been studied at higher level based on the ODE/IM correspondence for the third order ODE \cite{Bazhanov:2001xm}.
See also \cite{Bazhanov:2003ua,Ashok:2024zmw,Ashok:2024ygp,Ito:2024kza,Kudrna:2025bzg,Fateev1988} for
$W_N$ algebra, which is the algebra associated with the affine Lie algebra $A_{N-1}^{(1)}$. For a general affine Lie algebra, the ODE takes the form of a system of first-order linear differential equations, which can be obtained from the linear problem for the affine Toda field equations in the conformal/light-cone limit \cite{Lukyanov:2010rn,Ito:2013aea}. The WKB expansion of the linear system for the classical affine Lie algebras has been studied in \cite{Ito:2023zdc}, where the WKB coefficients are the same as the conserved currents in the Drinfeld-Sokolov reduction of the soliton equations hierarchy \cite{Drinfeld:1984qv}. Moreover, in \cite{Ito:2024kza}, the integrals of motion are calculated for $WA_r$ and $WD_r$ algebras \cite{Lukyanov:1989gg}, and the relation to the WKB expansions is confirmed at higher order level.

We will explore the relation between the integrals of motion of the W-algebras and the WKB expansions of the linear problem for other affine Lie algebras. In particular, exceptional type affine Lie algebras and twisted affine Lie algebras are interesting since the corresponding W-algebra has not been well studied so far. In this paper, we study the WKB expansion of the $E_6^{(1)}$-type affine Lie algebra and its relation to the integrals of motion of the $WE_6$-algebra, since this example is the simplest non-trivial Lie algebra, whose W-algebra is known \cite{Keller:2011ek}. Other types of affine Lie algebras will be studied in separate papers.

The representation of the Lie algebra $E_6$ is quite complicated since its nontrivial lowest dimension is 27. It is impossible to rewrite the linear problem into the form of a single ODE, which is useful for study of the $A_r$-type linear problem. We will explore the diagonalization method to study the WKB solution of the linear problem. The $E_6$ example provides a nontrivial test to this method. Unlike the $WA_r$ and $WD_r$ algebras, the generators of the $WE_6$ algebra are not constructed by the method of quantum Miura transformation, which has a close relationship with the soliton hierarchy. The ODE/IM correspondence for the $E$-type affine Lie algebra provides another nontrivial link between classical and quantum integrable models.

This paper is organized as follows. In Section 2, we first explain the basic properties of the Lie algebra $\mathfrak{g}$ and the affine Lie algebra $\hat{\mathfrak{g}}$. Then, we define the linear differential equation associated with the affine Lie algebra $\hat{\mathfrak g}$. Next, we discuss the WKB expansion of the solution to the linear problem, solved by diagonalizing the connection. In Section 3, we apply the method introduced in Section 2 to the $E_6^{(1)}$. We obtain the Riccati equations and solve them recursively to find the WKB solution to the $E_6^{(1)}$-type linear problem up to the sixth order.
We then compute their period integrals. In Section 4, we calculate the integrals of motion in the $WE_6$ CFT up to spin-6. They are shown to agree with the integrals. This provides strong evidence for the ODE/IM correspondence for $E_6^{(1)}$.

\section{The linear  problem for affine Lie algebra and the WKB solution}
In this section, we first summarize the basic properties of the Lie algebra and the conventions used in the present paper.
Let $\mathfrak{g}$ be a simple Lie algebra of rank $r$ and $\{E_\alpha, H^i\}$ ($\alpha \in \Delta$, $i=1,\dots,r$) its generators, where
$\Delta$ is the set of roots. 
The commutation relations for the generators are defined by
\begin{align}
[H^i,H^j]&=0,\\
[H^i,E_\alpha]&=\alpha^i E_\alpha,\\
[E_\alpha,E_\beta]&=
\left\{
\begin{array}{cc}
N_{\alpha,\beta}E_{\alpha+\beta}, & \text{for $\alpha+\beta \in \Delta$},\\
\alpha^{\vee}\cdot H, & \text{for $\alpha+\beta=0$},\\
0, &\mbox{otherwise},
\end{array}
\right.
\end{align}
where $\alpha\cdot H=\sum_{a=1}^{r}\alpha^a H^a$. $\alpha^{\vee}={2\alpha\over \alpha^2}$ is the coroot of $\alpha$.
$N_{\alpha,\beta}$ are the structure constants.
Let $\alpha_i$ and $\omega_i$ ($i=1,\dots, r$) be the simple roots and the fundamental weights, respectively. They satisfy
$\omega_i\cdot \alpha_j=\delta_{ij}$. The Cartan matrix is defined as $K_{ij}=\alpha_i\cdot \alpha^{\vee}_j$.
The (co-)Weyl vector $\rho$ ($\rho^{\vee}$) is the sum of (co-)fundamental weights.

Denote $\hat{\mathfrak g}={\mathfrak g}^{(\ell)}$ an affine Lie algebra associated with a simple Lie algebra ${\mathfrak g}$.
The index $\ell=1,2,3$ labels the degree of twist of the affine Lie algebra.
The structure of the affine Lie algebra is characterized by 
the extended root $\alpha_0$. For the case $\ell=1$, $\alpha_0=-\theta$, where $\theta$ is the highest root.
The (co)labels $n_i$ ($n^{\vee}_i$) are defined as integers that satisfy $\sum_{i=0}^r n_i\alpha_i=\sum_{i=0}^{r}n^{\vee}_i \alpha^{\vee}_i=0$ normalized 
to $n_0=n_0^{\vee}=1$. The (dual) Coxeter number $h$ ($h^{\vee}$) is given by the sum of the (co)labels.
$\hat{\mathfrak g}^{\vee}$ denotes the Langlands dual of $\hat{\mathfrak g}$, whose simple roots are $\alpha_i^{\vee}$.
In particular, simply-laced affine Lie algebras $A_r^{(1)}$, $D_r^{(1)}$, and $E_{6,7,8}^{(1)}$, whose squared norms of simple roots are two, are self-dual.

We now present the system of linear differential equations associated with an affine Lie algebra $\hat{\mathfrak g}$, which is obtained from the light-cone and the conformal limit of those for the affine Toda field equation modified by the conformal transformation specified by a holomorphic function $p(z)$ \cite{Lukyanov:2010rn,Ito:2013aea}.
For a representation $V$ of ${\mathfrak g}$, we define the linear differential equation for the $V$-valued function $\Psi(z)$ of a complex variable $z$ \cite{Sun:2012xw}:
\begin{align}
\Bigl(\epsilon \partial_z +A(z) \Bigr)\Psi(z)&=0,
\label{eq:lp1}
\end{align}
where $A(z)$ is the gauge connection defined by
\begin{align}
A(z)&=\epsilon \sum_{i=1}^{r}v^i(z)\alpha^{\vee}_i\cdot H+\sum_{i=1}^{r}E_{\alpha_i} +p(z)E_{\alpha_0}
\label{eq:ll1}
\end{align}
with
\begin{align}
 v^i(z)&={l_i\over z},~~~i=1,\cdots,r.
 \label{vi}
\end{align}
Here, $l_i$ are real parameters.
$\epsilon$ is a complex parameter that plays the role of the Planck constant in the WKB expansion. 
$p(z)$ is a polynomial in $z$. In this paper, we consider the case where $p(z)$ is a monomial in $z$ of the form:
\begin{align}
 p(z)&=z^{hM}-1.
 \label{pz}
\end{align}
Here $h$ is the Coxeter number of ${\mathfrak g}$, and $M$ is a positive real number.

We study the WKB solution of the linear problem \eqref{eq:lp1}.
A way to obtain the WKB expansion is to find the Riccati equation, from which one can derive the recursive relations for the WKB coefficients. For the $A_r^{(1)}$-type linear problem in the fundamental representation, one can find the higher-order derivative generalization of the Schr\"odinger equation. The Riccati equation can be easily generalized \cite{Ito:2021boh,Ito:2021sjo}. However, for the $D_r^{(1)}$ and $E_r^{(1)}$ types, it is difficult to apply this approach, as it is necessary to introduce the pseudo-differential operator $\partial^{-1}$ to obtain the single ODE for the highest weight component in $\Psi$.

We employ a different approach in \cite{Ito:2023zdc}.
The linear problem  \eqref{eq:lp1} can be transformed by the gauge transformation:
\begin{align}
A^g(z)&=g^{-1}(z)A(z)g(z)+\epsilon g^{-1}(z)\partial_z g(z),
\label{eq:gauge1}\\
\Psi^g(z)&=g(z)^{-1}\Psi(z),
\label{eq:gauge2}
\end{align}
where $g(z)\in G$, and $G$ is the Lie group of ${\mathfrak g}$.
Once we diagonalize the connection  $A(z)$ by a gauge transformation, the WKB solution can be found immediately.
In \cite{Ito:2023zdc}, it is found that the constraints for the gauge parameters reduce to the Riccati equation of the higher-order ODE for the $A$-type.
Moreover, it is applied to the WKB expansion for $D$-type and other classical non-simply laced affine Lie algebras.
The WKB expansion, where $\epsilon$ is the expansion parameter, defines the classical integrals of motion
for the integrable equations of Drinfeld-Sokolov \cite{Drinfeld:1984qv}.
Then, the WKB series of the solutions represents the classical integrals of motion.

\subsection{Gauge transformation and the Riccati equations}
Let us discuss the procedure for obtaining the Riccati equation for the linear problem associated with $\hat{\mathfrak g}$ by diagonalization. $A(z)$ can be represented by an $n\times n$ matrix, where $n$ denotes the dimension of the representation.
We consider the diagonalization of $A(z)$ by gauge transformation \eqref{eq:gauge1}, where $g(z)$ is given by
\begin{align}
g(z)&=\sum_{i=1}^{n}E_{ii}+\sum_{i=1}^{n-1}g_{ni}(z)E_{ni},
\label{gtm}
\end{align}
where $E_{ab}$ is the matrix whose $(i,j)$ entry is $\delta_{ia}\delta_{jb}$.
After the gauge transformation, the components in the bottom row of $A^g$ are found to be
\begin{align}
(A^g)_{ni}&=\begin{cases}
A_{ni}-\sum_{j=1}^{n-1}g_{nj}A_{ji}+A_{nn}g_{ni}-g_{ni}\sum_{j=1}^{n-1}g_{nj}A_{ji}+\epsilon \partial_z g_{ni}~~ (i\leq n-1),\\
A_{nn}-\sum_{j=1}^{n-1}g_{nj}A_{jn}~~(i=n).
\end{cases}
\end{align}
The bottom row of the diagonalized $A^g$ implies that the $(n-1)$ gauge parameters $g_{ni}$ should satisfy the equations
\begin{align}
(A^g)_{n1}&=\cdots =(A^g)_{nn-1}=0.
\label{eq:ric1}
\end{align}
We call Eqs. \eqref{eq:ric1} the Riccati equations for the linear problem \eqref{eq:lp1}. These equations are the non-linear quadratic equations for $g_{ni}$.
When we expand $g_{ni}(z)$ in $\epsilon$ as
\begin{align}
g_{ni}(z)&=\sum_{k=0}^{\infty}\epsilon^k s_{i}^{(k)}(z),
\label{eq:wkb2}
\end{align}
and substitute this into \eqref{eq:ric1}, $s_{i}^{(k)}(z)$ is determined recursively. 
For the zeroth order in $\epsilon$, the Riccati equations \eqref{eq:ric1} read
\begin{align}
(A^{(0)})_{ni}-\sum_{j=1}^{n-1}(s_{j}^{(0)}-s_{i}^{(0)} s_{j}^{(0)})(A^{(0)})_{ji}=0, \quad i=1,\ldots, n-1,
\label{eq:ric0}
\end{align}
where $A^{(0)}$ is the zeroth order term in the connection, which is given by
\begin{align}
A^{(0)}&=\sum_{i=1}^{r}E_{\alpha_i} +p(z)E_{\alpha_0}.
\end{align}
Eqs. \eqref{eq:ric0} are solved for $s_{i}^{(0)}$ ($i=0,\dots,n-1$) in terms of $p(z)$.
For higher order terms in $\epsilon$, we observe that the Riccati equations \eqref{eq:ric1} are quadratic in the gauge parameters $g$ as in the case of the Schr\"odinger equation. 
Then, in the $\epsilon^k$ term, the coefficients of $s^{(k)}_i$ ($i=0,\dots,n-1$)
are expressed as linear functions of $s_i^{(0)}$. The coefficients of order $\epsilon^k$ in the Riccati equations can be written in matrix form:
\begin{align}
B S_k
&=J_k, \quad S_k=\begin{pmatrix}
s_{0}^{(k)}\\
\vdots\\
s_{n-1}^{(k)}
\end{pmatrix},
\end{align}
where $B$ is the $(n-1)\times (n-1)$ matrix defined by
\begin{align}
B_{ij}&=\left. {\partial (A^g)_{ni} \over \partial g_{n,j}}\right|_{\epsilon=0,g_{n,l}=s_{l}^{(0)}}, \quad i,j=1,\dots,n-1,
\end{align}
and $J_k$ is the $(n-1)$ vector containing the lower order terms.
Then $S_k$ is determined as $B^{-1}J_k$, from which we can solve the WKB expansion of the solution of the diagonalized linear problem.
Since the Weyl transformation of the solution exchanges the components of $\Psi^g$ in \eqref{eq:gauge2}, we observe that the Riccati equations contain the full information of the 
WKB solutions.

\begin{figure}[h]
\centering
\includegraphics[width=6cm]{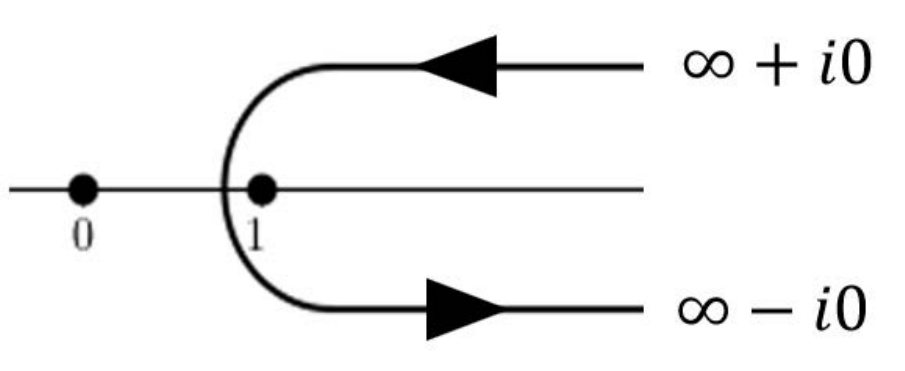}
  \caption{The Pochhammer contour.}
  \label{pochhammer}
\end{figure}

Practically, in some low-dimensional representation examples, we can take 
$A_{jn}=\delta_{jn-1}$. 
The lowest component of the diagonalized connection is given by
\begin{align}
(A^g)_{n,n}&=\epsilon~(\sum_{i=1}^{r}v^i(z) H_i)_{n,n}-g_{n,n-1}(z).
\label{Agnn}
\end{align}
Then, to obtain the WKB expansion, we need to find $g_{n,n-1}(z)=\sum_{k=0}^{\infty}\epsilon^k s_{n-1}^{(k)}(z)$.
$p(z)$ is expressed in terms of $s_{n-1}^{(0)}(z)$ as
\begin{align}
p(z)&=t [s_{n-1}^{(0)}(z)]^h
\end{align}
for some constant $t$ where $h$ is the Coxeter number of ${\mathfrak g}$.
The WKB solution is now given by
\begin{align}
(\Psi^g)_n&=\exp\left(-{1\over \epsilon}\int^z dx (A^g)_{n,n}(x)\right).
\end{align}
The WKB periods defined by the integral over a cycle $C$ on the complex plane are
expanded as
\begin{align}
    -\oint_C dz (A^g)_{n,n}(z)&=\sum_{k=0}^{\infty}\epsilon^k Q_k,
\end{align}
where
\begin{align}
Q_k&=\oint_C dz\Bigl(s^{(k)}_{n-1}(z)-\delta_{k,1}(\sum_{i=1}^rv^i(z)H_i)_{n,n}\Bigr).
\label{defWKBperiod}
\end{align}
In this paper, we take $C$ as the Pochhammer contour that starts from $z=\infty+i0$, goes just above the real axis, turns around $z=1$ in a half-turn, and goes just below the real axis to end at $z=\infty-i0$ \cite{Babenko:2017fmu} (See Figure \ref{pochhammer}). We call (\ref{defWKBperiod}) the $k$-th period.

We will discuss the relation between $Q_k$ and the integrals of motion in CFT.
For classical affine Lie algebras with low ranks, the diagonalization procedure mentioned above has been studied in \cite{Ito:2023zdc}. In the next section, we will apply the method to the exceptional affine Lie algebra $E_6^{(1)}$, where the representation is high-dimensional.

\setcounter{equation}{0}
\section{Linear differential equation and the WKB period for $E_6^{(1)}$}

We consider the 27-dimensional representation of the simply-laced Lie algebra $E_6$. The generators for the simple roots $\alpha_1,\cdots,\alpha_6$ are explicitly given by
\cite{Ito:2021boh}:
\begin{align}
E_{\alpha_1}
&=
E_{1,2}+E_{12,15}+E_{14,17}+E_{16,19}+E_{18,21}+E_{20,22},
\nonumber\\
E_{\alpha_2}
&=
E_{2,3}+E_{10,12}+E_{11,14}+E_{13,16}+E_{21,23}+E_{22,24},
\nonumber\\
E_{\alpha_3}
&=
E_{3,4}+E_{8,10}+E_{9,11}+E_{16,18}+E_{19,21}+E_{24,25},
\nonumber\\
E_{\alpha_4}
&=
E_{4,5}+E_{6,8}+E_{11,13}+E_{14,16}+E_{17,19}+E_{25,26},
\nonumber\\
E_{\alpha_5}
&=
E_{5,7}+E_{8,9}+E_{10,11}+E_{12,14}+E_{15,17}+E_{26,27},
\nonumber\\
E_{\alpha_6}
&=
E_{4,6}+E_{5,8}+E_{7,9}+E_{18,20}+E_{21,22}+E_{23,24},
\nonumber\\
E_{\alpha_0}
&=
E_{20,1}+E_{22,2}+E_{24,3}+E_{25,4}+E_{26,5}+E_{27,7},
\label{eq:e6-1}
\end{align}
where $E_{a,b}(a,b=1,...,27)$ is the 27-dimensional matrix whose $(k,l)$ entry is $\delta_{ak}\delta_{bl}$. We define $E_{-\alpha_i}:={}^tE_{\alpha_i}$ and
$H_i:=\alpha_i\cdot H=\comm{E_{\alpha_i}}{E_{-\alpha_i}}$. The explicit forms of $H_i$'s are as follows:
{\small
\begin{align}
  H_1
  &=
  E_{1,1}-E_{2,2}+E_{12,12}+E_{14,14}-E_{15,15}+E_{16,16}-E_{17,17}+E_{18,18}-E_{19,19}+E_{20,20}-E_{21,21}-E_{22,22},
  \nonumber\\
  H_2
  &=
  E_{2,2}-E_{3,3}+E_{10,10}+E_{11,11}-E_{12,12}+E_{13,13}-E_{14,14}-E_{16,16}+E_{21,21}+E_{22,22}-E_{23,23}-E_{24,24},
  \nonumber\\
  H_3
  &=
  E_{3,3}-E_{4,4}+E_{8,8}+E_{9,9}-E_{10,10}+E_{11,11}+E_{16,16}-E_{18,18}+E_{19,19}-E_{21,21}+E_{24,24}-E_{25,25},
  \nonumber\\
  H_4
  &=
  E_{4,4}-E_{5,5}+E_{6,6}-E_{8,8}+E_{11,11}-E_{13,13}+E_{14,14}-E_{16,16}+E_{17,17}-E_{19,19}+E_{25,25}-E_{26,26},
  \nonumber\\
  H_5
  &=
  E_{5,5}-E_{7,7}+E_{8,8}-E_{9,9}+E_{10,10}-E_{11,11}+E_{12,12}-E_{14,14}+E_{15,15}-E_{17,17}+E_{26,26}-E_{27,27},
  \nonumber\\
  H_6
  &=
  E_{4,4}+E_{5,5}-E_{6,6}+E_{7,7}-E_{8,8}-E_{9,9}+E_{18,18}-E_{20,20}+E_{21,21}-E_{22,22}+E_{23,23}-E_{24,24},
  \nonumber\\
  H_0
  &=
  -E_{1,1}-E_{2,2}-E_{3,3}-E_{4,4}-E_{5,5}-E_{7,7}+E_{20,20}+E_{22,22}+E_{24,24}+E_{25,25}+E_{26,26}+E_{27,27}.
  \label{eq:e6-2}
\end{align}
}The Cartan matrix is given by $K_{ij}={1\over 6}{\rm tr}H_i H_j$. The Coxeter number is $h=12$.

We consider the linear problem for $E_6^{(1)}$. The gauge connection (\ref{eq:ll1}) is now in the form:
\begin{align}
A(z)&=\epsilon \sum_{i=1}^{6}v^i(z)H_i+\sum_{i=1}^{6}E_{\alpha_i} +p(z)E_{\alpha_0}.
\label{eq:vandp}
\end{align}
$v^i(z)$ and $p(z)$ are given in \eqref{vi},\eqref{pz}. The gauge transformation matrix (\ref{gtm}) now takes the form:
\begin{align}
g(z)&=\sum_{i=1}^{27}E_{ii}+\sum_{i=1}^{26}g_{27,i}(z)E_{27,i}.
\end{align}
By the diagonalization procedure, we obtain the 26 Riccati equations:
\begin{align}
(A^g)_{27,1}&=\cdots =(A^g)_{27,26}=0,
\label{ag=0}
\end{align}
whose explicit forms are shown in Appendix \ref{app:ric1}. These equations determine the 26 gauge parameters $g_{27,i}(z)(i=1,...,26)$. By expanding $g_{27,i}(z)$ and $(A^g)_{27,i}$ as
\begin{align}
g_{27,i}(z)&=\sum_{k=0}^{\infty}\epsilon^ks_{i}^{(k)}(z),~~~~~(A^g)_{27,i}(z)=\sum_{k=0}^{\infty}\epsilon^kA_{i}^{(k)}(z),
\label{gA}
\end{align}
and substituting these into the equations \eqref{ag=0}, we can solve the resulting equations order by order. One can then obtain the coefficients $s_{i}^{(k)}(z)$ recursively. Let us now solve the Riccati equations (\ref{ag=0}) order by order. The 0-th order equations are given by
\begin{align}
A_{1}^{(0)}(z)&=\cdots =A_{26}^{(0)}(z)=0,
\end{align}
which are homogeneous for $s_{1}^{(0)}(z),...,s_{26}^{(0)}(z)$, and $p(z)$.
We aim to express $s_{1}^{(0)}(z),...,s_{25}^{(0)}(z)$, and $p(z)$ in terms of $s_{26}^{(0)}(z)$. To do so, we first assume that
\begin{align}
p(z)&=t [s_{26}^{(0)}(z)]^h
\label{ans}
\end{align}
with $h=12$ and the constant $t$ to be fixed. The 25 equations $A_{2}^{(0)}(z)=\cdots =A_{26}^{(0)}(z)=0$ determine the 25 functions $s_{1}^{(0)}(z),\cdots,s_{25}^{(0)}(z)$ in terms of $t$ and $s_{26}^{(0)}(z)$. For example, we find
\begin{align}
s_{1}^{(0)}(z)&= \frac{1}{117} \left(-1+252t\right) (s_{26}^{(0)}(z))^{16},~~~s_{20}^{(0)}(z)={1\over 13}(2+3t) (s_{26}^{(0)}(z))^{5}.
\end{align}
All the solutions are shown in \eqref{eq:zeroth1}. 
The parameter $t$ is determined by solving
\begin{align}
0=A_{1}^{(0)}(z)=-p(z)s_{20}^{(0)}(z)-s_{1}^{(0)}(z)s_{26}^{(0)}(z)=-\frac{1}{78} \left(27 t^2+270 t-1\right) (s_{26}^{(0)}(z))^{17}
\end{align}
as
\begin{align}
\label{valueoft}
t=\frac{1}{9}\qty(-45\pm26\sqrt{3}).
\end{align}
Then, one obtains $s_{26}^{(0)}(z)$ by (\ref{ans}). 
Because $s_{1}^{(0)}(z),\cdots,s_{25}^{(0)}(z)$, and $p(z)$ are expressed in terms of $s_{26}^{(0)}(z)$, there are 24 independent solutions to the 0-th order equations depending on the value of $t$ and the choice of the 12-th roots in (\ref{ans}).

The $k$-th order $(k\ge1)$ Riccati equation is given by $R_k=0$ with the vector:
\begin{align}
R_k:=\mqty(A_{1}^{(k)}(z) \\ \vdots \\ A_{26}^{(k)}(z)).
\end{align}
By substituting $g_{27,i}(z)$ in \eqref{gA}, $R_k$ turns out to be of the form:
\begin{align}
R_k=BS_k-J_k,~~~S_k=\mqty(s_{1}^{(k)}(z)\\
\vdots\\
s_{26}^{(k)}(z)).
\label{eq:rice6_2}
\end{align}
$B$ is the $26\times26$ matrix whose $(i,j)$ entry is
\begin{align}
B_{ij}&=\left.\frac{\partial (A^g)_{27,i}}{\partial g_{27,j}}\right|_{\epsilon=0,g_{27,l}=s_{l}^{(0)}}.
\end{align}
Explicitly,
\begin{align}
B&=E_{2,1}+E_{3,2}+E_{4,3}+E_{5,4}+E_{6,4}+E_{7,5}+E_{8,5}+E_{8,6}+E_{9,7}+E_{9,8}+E_{10,8}
\nonumber\\
&+E_{11,9}+E_{11,10}
+E_{12,10}+E_{13,11}+E_{14,11}+E_{14,12}+E_{15,12}
+E_{16,13}+E_{16,14}
\nonumber\\
&
+E_{17,14}+E_{17,15}+E_{18,16}+E_{19,16}+E_{19,17}+E_{20,18}+E_{21,18}+E_{21,19}
\nonumber\\
&
+E_{22,20}+E_{22,21}+E_{23,21}+E_{24,22}+E_{24,23}+E_{25,24}+E_{26,25}
+s_{26}^{(0)} I_{26}
\nonumber\\
&
+p(x)(E_{1,20}+E_{2,22}+E_{3,24}+E_{4,25}+E_{5,26})
+\sum_{i=1}^{26} s_{i}^{(0)}E_{i,26},
\end{align}
where $I_{26}$ is the identity matrix. $J_k$ in \eqref{eq:rice6_2} contains only lower order functions $s_{i}^{(j)}(z)~(j<k)$, which have already been determined in the former steps. One obtains $S_k$ from the $k$-th order Riccati equations $R_k=0$ as
\begin{align}
S_k=B^{-1}J_k.
\end{align}

The WKB coefficients $s_{26}^{(k)}~(k\ge1)$ are expressed in terms of $s_{26}^{(0)}(z)$ (or $p(z)$) and $l_i$ in \eqref{eq:vandp}. The coefficients are concisely written with the Casimirs of $E_6$ that we define as
\begin{align}
C_i&={1\over12}{\rm tr}(q\cdot H)^i, \quad i=2,3,\cdots.
\end{align}
Here, $q=l+\rho$, $l=\sum_{i=1}^{6}l_i\alpha_i$, and $\rho$ is the Weyl vector:
\begin{align}
\rho&=8\alpha_1+15\alpha_2+21\alpha_3+15\alpha_4+8\alpha_5+11\alpha_6.
\end{align}
The independent elements are $C_2$, $C_5$, $C_6$, $C_8$, $C_9$ and $C_{12}$. One finds
\begin{align}
C_2&={1\over2}q_i K_{ij}q_j,~~C_3=0,~~C_4=C_2^2,~~C_7={7\over2}C_5 C_2,~~\cdots.
\end{align}
From the Riccati equations $R_k=0~(k=1,2,3,4)$, the coefficients $s_{26}^{(k)}$ are obtained as
\begin{align}
s_{26}^{(1)}&=-v^5(z)+8{(s_{26}^{(0)})' \over s_{26}^{(0)}},\\
\label{s262}
s_{26}^{(2)}&={3\pm \sqrt{3} \over12}\left({C_2-39\over 2z^2(s_{26}^{(0)})^2}-117 {((s_{26}^{(0)})')^2 \over (s_{26}^{(0)})^3}+78 {(s_{26}^{(0)})'' \over (s_{26}^{(0)})^2}\right),\\
s_{26}^{(3)}&=-\qty(1\pm\frac{\sqrt{3}}{2})\qty(\frac{C_2-39}{z^3(s_{26}^{(0)})^2}+\frac{(C_2-39)(s_{26}^{(0)})'}{z^2(s_{26}^{(0)})^3}-39\frac{(s_{26}^{(0)})'''}{(s_{26}^{(0)})^3}+234\frac{(s_{26}^{(0)})'(s_{26}^{(0)})''}{(s_{26}^{(0)})^4}-234\frac{((s_{26}^{(0)})')^3}{(s_{26}^{(0)})^5}),\\
s_{26}^{(4)}&=0.
\end{align}
The sign $\pm$ depends on the double sign in $t$ in \eqref{valueoft}. Since we are interested in the period integrals over the closed contour, we can extract the terms that include $(s_{26}^{(0)})'$ by partial integration. This procedure reduces the number of terms in the integrands and gives the following coefficient $s_{26}^{(2)}$:
\begin{align}
\label{s262a}
s_{26}^{(2)}&=\frac{3\pm\sqrt{3}}{12}\qty(\frac{C_2-39}{z^2s_{26}^{(0)}}+\frac{39}{2}\frac{(s_{26}^{(0)})''}{(s_{26}^{(0)})^2})+\partial(\ast),
\end{align}
where $\partial(\ast)$ denotes the total derivative terms. From the Riccati equations $R_k=0~(k=5,6)$, $s_{26}^{(5)}$ and $s_{26}^{(6)}$ are obtained.
Up to the total derivative terms, they become\footnote{The Wolfram Mathematica program is attached as an ancillary file.}:
\begin{align}
s_{26}^{(5)}&=-\frac{3\pm2 \sqrt{3}}{30}~\frac{C_5}{z^5(s_{26}^{(0)})^4}+\partial(\ast),
\end{align}
\begin{align}
s_{26}^{(6)}&=\frac{176137 \left(2\pm\sqrt{3}\right) }{512 }\frac{(s_{26}^{(0)''})^3}{(s_{26}^{(0)})^8}+\frac{20485(2\pm\sqrt{3})}{1792}(C_2-39)\frac{(s_{26}^{(0)''})^2}{z^2(s_{26}^{(0)})^7}
\nonumber\\
&\quad
-\frac{29835 \left(2\pm\sqrt{3}\right)}{512}\frac{s_{26}^{(0)''}s_{26}^{(0)''''}}{(s_{26}^{(0)})^7}+\frac{85(2\pm\sqrt{3})}{1152}(2C_2^2-237C_2+6201)\frac{s_{26}^{(0)''}}{z^4(s_{26}^{(0)})^6}
\nonumber\\
&\quad
-\frac{11271 \left(2\pm\sqrt{3}\right)}{448}\frac{(s_{26}^{(0)'''})^2}{(s_{26}^{(0)})^7}+\frac{1445(2\pm\sqrt{3})}{672}(C_2-39)\frac{s_{26}^{(0)'''}}{z^3(s_{26}^{(0)})^6}
  \nonumber\\
  &\quad
  -\frac{9979(2\pm\sqrt{3})}{10752}(C_2-39)\frac{s_{26}^{(0)''''}}{z^2(s_{26}^{(0)})^6}+\frac{9945 \left(2\pm\sqrt{3}\right)}{7168}\frac{s_{26}^{(0)''''''}}{(s_{26}^{(0)})^6}
  \nonumber\\
\label{s2660328}
&\quad
+\frac{(2\pm\sqrt{3})}{4032}\left(168C_6-210C_2^3-1190C_2^2
  +98481C_2-2225691
  \right)
  \frac{1}{z^6(s_{26}^{(0)})^5}+\partial(\ast).
\end{align}

We now evaluate the period integrals $Q_k$ of the WKB coefficients $s_{26}^{(k)}$. For $E_6^{(1)}$, because $(\sum_{i=1}^6v^i(z)H_i)_{27,27}=-v^5(z)$, the $k$-th period (\ref{defWKBperiod}) now takes the form:
\begin{align}
\label{Qk}
Q_k&=\oint_C dz(s^{(k)}_{26}(z)+\delta_{k,1}v^5(z)).
\end{align}
Because the integrands in $Q_1$ and $Q_3$ are total derivatives, we obtain $Q_1=Q_3=0$. From $s_{26}^{(4)}=0$, we obtain $Q_4=0$.
We calculate $Q_2,Q_5$ and $Q_6$. Substituting $s_{26}^{(2)}$ in (\ref{s262a}) into (\ref{Qk}), $Q_2$ becomes
\begin{align}
\label{Q2A}
Q_2&=\frac{3\pm\sqrt{3}}{12}(C_2-39)\oint_C\dd z\frac{1}{z^2(s_{26}^{(0)})}+\frac{3\pm\sqrt{3}}{12}~\frac{39}{2}\oint_C\dd z\frac{s_{26}^{(0)''}}{(s_{26}^{(0)})^2}.
\end{align}
Substituting
\begin{align}
s_{26}^{(0)}=t^{-1/h}\qty(z^{hM}-1)^{1/h},
\label{s260}
\end{align}
which follows from (\ref{pz}) and (\ref{ans}), into the above two
integrals, they are shown to be written in terms of 
\begin{equation}
J(a,b):=\oint_Cdz(z^{hM}-1)^az^b=-\frac{2\pi ie^{i\pi a}}{hM}\frac{\Gamma(-a-\frac{b+1}{hM})}{\Gamma(-a)\Gamma(1-\frac{b+1}{hM})}.
\label{Jrec}
\end{equation}
We find
\begin{align}
\label{A1}
\oint_C\dd z\frac{1}{z^2(s_{26}^{(0)})}&=t^{1/h}J(-\frac{1}{h},-2),\\
\label{A2}
\oint_C\dd z\frac{s_{26}^{(0)''}}{(s_{26}^{(0)})^2}&=t^{1/h}M\qty((M-1)J(-\frac{1}{h}-1,-2+hM)+(1-h)MJ(-\frac{2}{h}-2,-2+hM)).
\end{align}
Using the recurrence relation for $J(a,b)$:
\begin{align}
\label{A3}
J\qty(a-m,b+n(hM))=J(a,b)e^{i\pi m}\frac{\Gamma\qty(\frac{b+1}{hM}+n)\Gamma(a-m+1)\Gamma\qty(a+1+\frac{b+1}{hM})}{\Gamma\qty(a+1+\frac{b+1}{hM}+n-m)\Gamma(a+1)\Gamma\qty(\frac{b+1}{hM})}
\end{align}
with $m,n\in\mathbb{Z}$, we can express the integrals \eqref{A1} and \eqref{A2} in terms of $J(-{1\over h},-2)$.
Finally, we obtain
\begin{align}
Q_2=t^{1/h}\frac{3\pm\sqrt{3}}{12}J(-\frac{1}{h},-2)\qty(C_2-36(M+1)).
\label{Q2inC2}
\end{align}

The fifth order period:
\begin{align}
Q_5=-\oint_C\dd z\frac{3\pm2 \sqrt{3}}{30}\frac{C_5}{z^5(s_{26}^{(0)})^4}
\label{Q5inC5}
\end{align}
is written simply in terms of $J(a,b)$ as
\begin{align}
Q_5=-t^{4/h}\frac{3\pm2\sqrt{3}}{30}J(-\frac{4}{h},-5)C_5.
\label{Q5inC5-2}
\end{align}

Finally, we compute the sixth-order correction to the period $Q_6$. Substituting $s_{26}^{(6)}$ in (\ref{s2660328}) into (\ref{Qk}), we get the sum of the nine contour integrals. Each of the integrals has the factor $t^{5/h}$. After expressing the integrals in terms of $J(-\frac{5}{h}-m,-6+n(hM))$ and using the recurrence relation (\ref{A3}), the integrals are factorized by $J(-\frac{5}{h},-6)$. The explicit form of every integral is shown in Appendix B. Using these formulae, we obtain
\begin{align}
\nonumber
Q_6&=t^{5/h}\frac{2\pm\sqrt{3}}{96}J(-\frac{5}{h},-6)\left[4C_6-5C_2^3-60(M+1)C_2^2-432(M+1)(24M^2-13M-13)C_2\right.\\
&\qquad\left.-5184(M+1)(288M^4-120M^3-95M^2+50M+25)\right].
\label{Q6inC6}
\end{align}
We have obtained the WKB expansion of the periods up to the sixth order. We can extend this calculation to higher orders. We find that the seventh-order period is zero. The eighth-order period is currently difficult to calculate.

\section{$WE_6$ algebra and integrals of motion}
In this section, we study the integrals of motion on a cylinder in two-dimensional $WE_6$ conformal field theory and compute their eigenvalues for the highest-weight state of the W-algebra.

The $WE_6$-algebra is generated by the higher spin currents $W_s$ of the spins $s=2,5,6,8,9,$ and $12$.
The free field realization of $WE_6$ algebra was studied in
\cite{Keller:2011ek}.
To construct them, we focus on the $A_5$ subalgebra of $E_6$. The associated $WA_5$ algebra has five generators, which are denoted by $w_k~(k=2,3,4,5,6)$. $w_k$ are expressed by free fields through the quantum Miura transformation \cite{Fateev1988}:
\begin{eqnarray}
\qty(a\partial_u)^6-\sum_{k=2}^6w_k(u)\qty(a\partial_u)^{6-k}=~\qty(a\partial_u-i\epsilon_1\cdot\partial\varphi(u))\cdots \qty(a\partial_u-i\epsilon_6\cdot\partial\varphi(u))~,
\label{miuratransf}
\end{eqnarray}
where $u$ is the coordinate on the complex plane, $a$ is a parameter, $\epsilon_i~(i=1,2,\cdots,6)$ are the weight vectors of the fundamental representation defined by $\epsilon_i=\omega_i-\omega_{i-1}$ with $\omega_0=\omega_6=0$. $\varphi=\qty(\varphi_1,\varphi_2,\varphi_3,\varphi_4,\varphi_5)$ are the free bosons that satisfy the OPE:
\begin{eqnarray}
\varphi_i(u)\varphi_j(v)=-\delta_{ij}\mathrm{log}\qty(u-v)+\cdots.
\end{eqnarray}
We will introduce $p_i=i\epsilon_i\cdot\partial\varphi(u)$.
In \eqref{miuratransf}, the RHS should be understood as the normal ordered product on the complex plane.
The Dynkin diagram of $E_6$ is invariant under the ${\mathbb Z}_2$ outer-automorphism,
which also induces the ${\mathbb Z}_2$-symmetry of the diagram of $A_5$ as
$\epsilon_i\rightarrow -\epsilon_{7-i}$.
We define the basis of generators of the $WA_5$ algebra with definite
${\mathbb Z}_2$ parity as
\begin{align}
\nonumber
\tilde{w}_2&:=w_2,\\
\nonumber
\tilde{w}_3&:=w_3-2a\partial w_2,\\
\nonumber
\tilde{w}_4&:=w_4-\frac{3}{2}a\partial \tilde{w}_3,\\
\nonumber
\tilde{w}_5&:=w_5-a\partial \tilde{w}_4+a^3\partial^3w_2,\\
\tilde{w}_6&:=w_6-\frac{a}{2}\partial \tilde{w}_5+\frac{a^3}{4}\partial^3\tilde{w}_3.
\end{align}
$\tilde{w}_k$ transforms as $(-1)^k \tilde{w}_k$ under the ${\mathbb Z}_2$-automorphism.
The $WE_6$-algebra is constructed from $\tilde{w}_k$ and a free boson $\phi(u)$.
We also define $p(u)=i\partial\phi(u)$. Then, the W-currents $W_s$ are given by
\begin{align}
W_2&=w_2+\frac{1}{2}(pp)-\frac{11}{\sqrt{2}}a\partial p,\label{eq:spin2}\\
W_5&=\tilde{w}_5+\frac{1}{2}\qty(\tilde{w}_3(pp))+\frac{3a^2}{2}\partial^2\tilde{w}_3-\frac{3a}{\sqrt{2}}\qty(\partial\tilde{w}_3p)-\sqrt{2}a\qty(\tilde{w}_3\partial p),\label{eq:spin5} \\
W_6&=\frac{1}{12a^2-1}\qty{W_5W_5}_4,\\
W_8&=\qty{W_5W_5}_2,\\
W_9&=\qty{W_5W_6}_2,\\
W_{12}&=\qty{W_6W_8}_2,
\end{align}
where $(AB)(u)$ denotes the normal ordered product of fields $A(u)$ and $B(u)$ with conformal dimensions $h_A$ and $h_B$, respectively, and $\{AB\}_k$ are the coefficients in the OPEs:
\begin{align}
A(u)B(v)=\sum_{m=1}^{h_A+h_B}\frac{\qty{AB}_m(v)}{(u-v)^m}+\qty(AB)(v)+\cdots.
\end{align}
In terms of free fields, the spin 2 current $W_2(u)$ is expressed as
\begin{align}
\label{emten1}
W_2(u)=-\sum_{i<j}(p_ip_j)-a\sum_{i=1}^5(i-1)\partial p_i+\frac{1}{2}p^2-\frac{11}{\sqrt{2}}a\partial p,
\end{align}
which can be written in terms of $\phi=(\varphi_1,\dots,\varphi_5,\phi_6)$ as
\begin{align}
    W_2(u)&=-{1\over2} (\partial\phi)^2-i a\rho\cdot \partial^2\phi.
    \label{emten2}
\end{align}
Here, the Weyl vector $\rho$ of $E_6$ is decomposed into the sum of the Weyl vector $\rho(A_5)$ of $A_5$ and its orthogonal direction with the unit vector $e_6$ as
\begin{align}
    \rho&=\rho(A_5)+{11\over \sqrt{2}} e_6,
\end{align}
where $\rho(A_5)=\sum_{i=1}^{6}(6-i)\epsilon_i$.
$\rho$ is normalized as $\rho^2=78$.
From (\ref{emten2}), the central charge is obtained as 
\begin{align}
    c=6-12a^2\rho^2=6-936a^2.
\end{align}
The spin 5 current  $W_5$ in \eqref{eq:spin5} is primary, but $W_6,W_8, W_9$ and $W_{12}$ are not. For example, we can define the spin-6 primary field by
\begin{eqnarray}
\widetilde{W}_6:=W_6+x_3\qty(W_2\qty(W_2W_2))+x_4\qty(W_2\partial^2W_2)+x_5\qty(\partial W_2\partial W_2)+x_6\partial^4W_2
\end{eqnarray}
with
\begin{align}
\nonumber
x_3&=\frac{20(1-42a^2)(157-23616a^2)}{9(11-1872a^2)(55-3276a^2)},\\
\nonumber
x_4&=\frac{40(1-42a^2)(11-10467a^2+918216a^4)}{9(11-1872a^2)(55-3276a^2)},\\
\nonumber
x_5&=\frac{20(1-42a^2)(58-11379a^2+512460a^4)}{3(11-1872a^2)(55-3276a^2)},\\
x_6&=\frac{5(42a^2-1)(78848640a^6+2325024a^4-53964a^2+161)}{27(11-1872a^2)(55-3276a^2)}.
\end{align}
The representation of the $WE_6$ algebra is characterized by the primary field $V_{\lambda}=: e^{i \Lambda\cdot \varphi}e^{i q\phi}:$. A pair $(\Lambda,q)$ is related to the weight vectors of $E_6$ by $\lambda=\Lambda+q e_6$. We define the W-charges $\Delta_s$ $(s=2,5,6,8,9,12)$ for the primary field $V_{\lambda}$ by
\begin{align}
    W_s(u) V(v)&={\Delta_s V(v)\over (u-v)^s}+\mbox{lower order terms}.
\end{align}
$\widetilde{\Delta}_6$ is defined by the OPE of $\widetilde{W}_6$ and $V$ similarly. 
The corresponding highest weight state $|\Delta\rangle$ is the eigenstate of the zero modes of $(W_s)_0$ where $W_s(u)=\sum_{n=-\infty}^{\infty}(W_s)_n u^{-n-s}$:
\begin{align}
(W_s)_0 |\Delta\rangle&=\Delta_s |\Delta\rangle.
\end{align}

For the $WE_6$ algebra, $\Delta_2$ is given by
\begin{align}
    \Delta_2&={1\over2}\Lambda \cdot(\Lambda+2a\rho(A_5))+{1\over2}q^2+a{11\over \sqrt{2}}q,
\end{align}
which can be expressed as
\begin{align}
\Delta_2&={1\over2}\lambda\cdot(\lambda+2a\rho).
\end{align}
The higher-order W-charges can be expressed in terms of Casimirs 
\begin{align}
   D_k:= {\rm tr}(\mu\cdot H)^k, \quad k=2,3,\cdots
\end{align}
associated with $\mu=\lambda+a\rho$.
For example, we find that the W-charges $\Delta_2$, $\Delta_5$ and $\widetilde{\Delta}_6$ are
given by
\begin{align}
\nonumber
\Delta_2&={1\over12}D_2-39a^2,\\
\nonumber
\Delta_5&={1\over 60}D_5,\\
\label{delta256}
\widetilde{\Delta}_6&=\Delta_6+x_3 \Delta_6^{(3)}+x_4 \Delta_6^{(4)}+x_5 \Delta_6^{(5)}+x_6 \Delta_6^{(6)},
\end{align}
where
\begin{align}
    \Delta_6&=-{1\over9}D_6+{1\over1296}D_2^3+{1\over2432}(-20+910a^2) D_2^2+{1\over108}(-40+8880a^2-306180a^4)D_2
\nonumber\\
&+{130\over3}a^2 (4-654a^2+20463a^4),\nonumber\\
    \Delta_6^{(3)}&={1\over1728}D_2^3+{1\over432}(18-351a^2)D_2^2+{1\over 108}(72-4212a^2+41067a^4)D_2
    \nonumber\\
&-39a^2(-4+39a^2)(-2+39a^2),\nonumber\\
    \Delta_6^{(4)}&={1\over24}D_2^2+{1\over108}(216-4212a^2)D_2+234a^2(-4+39a^2),\nonumber\\
    \Delta_6^{(5)}&={1\over36}D_2^2+{1\over108}(54-2808a^2)D_2+234a^2 (-1+26a^2),\nonumber\\
    \Delta_6^{(6)}&=10D_2-4660a^2.
\end{align}

Now, we perform the conformal transformation $u=e^{z}$ with the coordinate $z=i\sigma+\tau$ on a cylinder with the space parameter $\sigma\in\left[0,2\pi\right)$ and the time parameter $\tau\in\mathbb{R}$.
We define the conserved current of spin-$k$: $j_k(u)$ on the complex plane as a linear combination of spin-$k$ operators constructed from the W-currents and their derivatives. We can also define the conserved currents $j_k(z)$ on the cylinder by the conformal transformation from $j_k(u)$. The conserved charges are given by
\begin{eqnarray}
\label{iomdef}
\hat{I}_k:=\int_0^{2\pi}\frac{d\sigma}{2\pi}j_k(z).
\end{eqnarray}
These satisfy the involution conditions $[\hat{I}_k,\hat{I}_l]=0$.
If a conserved current is absent for a certain spin, we do not have the integral of motion for that spin. The conserved currents up to spin-6 are found to be
\begin{align}
j_2(u)&=W_2,\nonumber\\
j_5(u)&=W_5,\nonumber\\
j_6(u)&=W_6+y_1 (W_2(W_2 W_2))+y_2 (\partial W_2 \partial W_2).
\end{align}
Here, $y_1$ and $y_2$ are constants that are determined by the involution condition. In fact, the condition $[\hat{I}_5,\hat{I}_6]=0$ determines $y_1$ and $y_2$ as $y_1={1\over3}$
and $y_2={1\over36}(7+156a^2)$.
Eq. (\ref{iomdef}) implies that  $\hat{I}_k$ is given by the zero mode of $j_k(z)$ on the cylinder denoted by $\qty(j_k)_0$. We apply the operators $\hat{I}_k$ on the highest weight state $\left|\Delta\right>$, which is characterized as the eigenstate of the W-operators with eigenvalue $\Delta_s$.
The state $|\Delta\rangle$ is also the eigenstate for $\hat{I}_k$ whose eigenvalue is denoted by  $I_k$:
\begin{eqnarray}
\hat{I}_k\left|\Delta\right>=I_k\left|\Delta\right>.
\end{eqnarray}
First, we compute $I_2$. By the conformal transformation $z=\log u$, $W_2(u)$ transforms to $W_2(z)$ as
\begin{align}
W_2(z)=u^2W_2(u)-\frac{c}{24}.
\end{align}
Then the zero mode $(j_2)_0$ is $(W_2)_0-{c\over24}$ and we find
\begin{align}
    I_2&=\Delta_2-\frac{c}{24}.
\end{align}
For $I_5$, it is simply given by $\Delta_5$ since $W_5(u)$ is a primary field:
\begin{align}
    I_5&=\Delta_5.
\end{align}
For the spin 6 conserved current $j_6$, which is not primary, it is convenient to
express it as
\begin{align}
j_6(z)&=\widetilde{W}_6(z)+(y_1-x_3): W_2 (:W_2 W_2:):(z)+(y_2-x_5+x_4):\partial W_2 \partial W_2:(z)+\partial(*).
\end{align}
Here, $:\ :$ shows the symbol of the normal ordered product on the cylinder.
The zero mode of the first term is given by $\widetilde{\Delta}_6$. The zero modes of the second and third terms are found in \cite{Bazhanov:1994ft,Dymarsky:2019iny, Novaes:2021vjh}. Then we obtain
\begin{align}
I_6&=\widetilde{\Delta}_6+(y_1-x_3) I_6^{(1)}+ (y_2-x_5+x_4) I_6^{(2)},
\label{eq:iom6}
\end{align}
where $I_6^{(1)}$ and $I_6^{(2)}$ are given by 
\begin{align}
\nonumber
I_6^{(1)}&:=(:W_2(:W_2 W_2:):)_0=\Delta_2^3-{c+4\over 8}\Delta_2^2+\Bigl( {c^2\over192}+{7c\over 160}+{1\over 15}\Bigr)\Delta_2
-\Bigl({c^3\over 13824}+{11c^2\over 11520}+{47c\over 15120}\Bigr),
\\
I_6^{(2)}&:=(:\partial W_2 \partial W_2:)_0={31c \over  30240}-{1\over 60}\Delta_2.
\end{align}
Using \eqref{delta256}, we can express the eigenvalues $I_2$, $I_5$, and $I_6$ in terms of Casimirs associated with $\mu$, which become
\begin{align}
\label{I2inD2}
I_2&=\frac{D_2}{12}-\frac{1}{4},\\
\label{I5inD5}
I_5&=\frac{D_5}{60},\\
I_6&=-{1\over9}D_6 +{5\over 5184}D_2^3+{5\over 5184}D_2^2
+{-13+24a^2\over 1728}D_2+{25-120a^2+288a^4\over 1728}.
\label{I6inD6}
\end{align}

Let us compare these eigenvalues $I_2,I_5,I_6$ with the period integrals \eqref{Q2inC2}, \eqref{Q5inC5-2}, and \eqref{Q6inC6} derived in the previous section. We discuss the correspondence between the second-order period $Q_2$ and the eigenvalue $I_2$ of the integral of motion. We rewrite $Q_2$ as
\begin{equation}
Q_2=t^{1/h}h(3\pm\sqrt{3})J(-\frac{1}{h},-2)(M+1)\qty(\frac{C_2}{12h(M+1)}-\frac{1}{4}).
\end{equation}
By comparing the coefficients of $\frac{1}{4}$ terms in $Q_2$ and $I_2$, and imposing the following relation:
\begin{align}
\frac{C_2}{h(M+1)}=D_2,
\label{C2D2}
\end{align}
$Q_2$ and $I_2$ in (\ref{I2inD2}) are equal up to an overall coefficient:
\begin{align}
Q_2=t^{1/h}h(3\pm\sqrt{3})J(-\frac{1}{h},-2)(M+1)I_2.
\end{align}
This implies the following relation between the ODE parameters $(l,M)$ and the IM parameters $(\lambda,a)$:
\begin{align}
        \frac{l+\rho}{h\sqrt{M+1}}=\lambda+a\rho.
\label{paramrel1}
\end{align}
This is the same relation as that of the $A_r^{(1)}$ and $D_r^{(1)}$ cases \cite{Ito:2024kza}. \eqref{paramrel1} leads to the following identity for Casimirs by $q=l+\rho$ and $\mu=\lambda+a\rho$:
\begin{align}
    \tr(q\cdot H)^k=h^{k}(M+1)^{\frac{k}{2}}\tr(\mu\cdot H)^k.
\label{CD}
\end{align}
Applying this relation, $Q_5$ in (\ref{Q5inC5-2}) corresponds to $I_5$ in (\ref{I5inD5}) as
\begin{align}
  Q_5&=-t^{4/h}h^4~2\qty(3\pm2\sqrt{3})J(-\frac{4}{h},-5)(M+1)^{5/2}I_5.
\end{align}
For the sixth order, $Q_6$ agrees with $I_6$ as
\begin{align}
  Q_6=-t^{5/h}h^5~\frac{3}{8}\qty(2\pm\sqrt{3})J(-\frac{5}{h},-6)(M+1)^3I_6,
\end{align}
if we impose another relation for the parameters:
\begin{equation}
    a^2=\frac{M^2}{M+1}.
\label{Ma}
\end{equation}
This is the same relation as the one needed for the higher order $Q_k$ and $I_k$ to agree in the $A_r^{(1)}$ and $D_r^{(1)}$ cases \cite{Ito:2024kza}.

Thus, we find the correspondence between the WKB periods $Q_k$ and the eigenvalues $I_k$ of the integrals of motion under the same parameter relations as those of $A_r^{(1)}$ and $D_r^{(1)}$ types.
Note that when $a$ is parametrized by $\beta$ as $a=\beta-{1\over \beta}$, $\beta$ is given by $\beta=\pm \sqrt{M+1}$.

We expect that the higher-order WKB periods and the eigenvalues of the integrals of motion match under the relations \eqref{paramrel1} and \eqref{Ma}. However, these are currently difficult to compute, and their comparison is left for future study.

\section{Conclusions and Discussion}

In this paper, we consider the $E_6^{(1)}$-type linear problem and obtain the WKB solution up to the sixth order. We compute their period integrals along the Pochhammer contour. Then, we calculate the integrals of motion in the CFT with the $WE_6$-algebra up to spin-6. These integrals of motion are shown to agree with the period integrals when the parameters satisfy the same relations as those in the $A_r^{(1)}$ and $D_r^{(1)}$ cases. Our result provides strong evidence for the ODE/IM correspondence for the exceptional type affine Lie algebra $E_6^{(1)}$.

It is interesting to study the WKB expansions for the other exceptional affine Lie algebras $E_7^{(1)}$  and $E_8^{(1)}$, where the structures of the corresponding W-algebras are not yet known. It is also interesting to study the WKB expansion for non-simply laced affine Lie algebras. The structure of $W\mathfrak{g}$ algebras is less known for non-simply laced $\mathfrak{g}=B_n^{(1)},C_n^{(1)}$, $F_4^{(1)}$, and $G_2^{(1)}$ \cite{Ito:1995ny}, and it is expected that they correspond to the WKB expansions for the Langlands dual of $\mathfrak{g}$, namely, $\mathfrak{g}^{\vee}=A_{2n-1}^{(2)},D_{n+1}^{(2)}$, $E_6^{(2)}$, and $D_4^{(3)}$, respectively \cite{deBoer:1993iz,Kausch:1991zm,Ito:2013aea}. Our approach will be useful for understanding its free field representation via the integrals of motion. In particular, we recover the eigenvalues of the normal ordered products of the generators of W-currents
from the WKB periods, which provide important information to reconstruct the W-algebra.
These W-algebras will be useful to understand the structure of the Nekrasov partition function for arbitrary gauge group \cite{Keller:2011ek} and the quantum Seiberg-Witten curve for Argyres-Douglas theories \cite{Ito:2017ypt,Ito:2024wxw}.

The WKB periods in the $A_r$-type SW theory, the $(A_2,A_N)$-type Argyres-Douglas theory are shown to give the thermodynamic Bethe ansatz (TBA) equations in the related integrable models
\cite{Ito:2021boh,Ito:2021sjo,Grassi:2019coc,Grassi:2021wpw, Fioravanti:2019awr,Fioravanti:2019vxi,Fioravanti:2022bqf}. These equations are shown to exhibit the wall-crossing phenomena in the strong-coupling dynamics. It is interesting to study the TBA equations and wall-crossing phenomena associated with the WKB periods for $E_6$ in the present paper.

\section*{Acknowledgments}
We would like to thank Mingshuo Zhu, Shigeki Miyazaki, and Naozumi Tanabe for their useful discussions and comments.
D.I. and W.K. are supported by the Tsubame Scholarship for Doctoral Students at Institute of Science Tokyo.

\appendix
\renewcommand{\theequation}{\Alph{section}.\arabic{equation}}
\setcounter{equation}{0}
\section{The Riccati equations for $E_6^{(1)}$}\label{app:ric1}
In this Appendix, we present the Riccati equations for the linear problem associated with $E_6^{(1)}$, which are
\begin{align}
(A^g)_{27,1}&=-p g_{20}-\epsilon (v_1+v_5) g_1-g_1 g_{26}+\epsilon g'_1=0,\nonumber\\
(A^g)_{27,2}&= -g_1-p g_{22} -\epsilon g_2(-v_1+v_2+v_5) -g_2 g_{26}+\epsilon g'_2=0,\nonumber\\
(A^g)_{27,3}&=-g_2-p g_{24}-\epsilon g_3 (-v_2+v_3+v_5)-g_3 g_{26}+\epsilon g'_3=0,\nonumber\\
(A^g)_{27,4}&=-g_3 -p g_{25} -\epsilon g_4 (v_5-v_3+v_4+v_6)-g_4 g_{26}+\epsilon g'_4=0,
\nonumber\\
(A^g)_{27,5}&=-g_4-p g_{26}-g_5 g_{26}-\epsilon g_5 (-v_4+2v_5+v_6)+\epsilon g'_5=0,
\nonumber\\
(A^g)_{27,6}&=-g_4-g_6 g_{26} -\epsilon g_6 (v_5+v_4-v_6)+\epsilon g'_6=0,\nonumber\\
(A^g)_{27,7}&=p-g_5-g_7 g_{26}-\epsilon g_7 v_6+\epsilon g'_7=0,\nonumber\\
(A^g)_{27,8}&=-g_5-g_6-g_8 g_{26}-\epsilon g_{8}(2v_5+v_3-v_4-v_6)+\epsilon g'_8=0,\nonumber\\
(A^g)_{27,9}&=-g_7-g_8 -g_9 g_{26} -\epsilon g_9(v_3-v_6)+\epsilon g'_9=0,\nonumber\\
(A^g)_{27,10}&=-g_8-g_{10}g_{26} -\epsilon g_{10} (2v_5+v_2-v_3)+\epsilon g'_{10}=0,\nonumber\\
(A^g)_{27,11}&=-g_9-g_{10} -g_{11}g_{26}-\epsilon g_{11}(v_2-v_3+v_4)+\epsilon g'_{11}=0,
\nonumber\\
(A^g)_{27,12}&=-g_{10}-g_{12}g_{26} -\epsilon g_{12} (2v_5+v_1-v_2)+\epsilon g'_{12}=0,\nonumber\\
(A^g)_{27,13}&=-g_{11}-g_{13}g_{26}-\epsilon g_{13} (v_2-v_4+v_5) +\epsilon g'_{13}=0,\nonumber\\
(A^g)_{27,14}&=-g_{11}-g_{12}-g_{14}g_{26}-\epsilon g_{14} (v_1-v_2+v_4)+\epsilon g'_{14}=0,\nonumber\\
(A^g)_{27,15}&=-g_{12}-g_{15}g_{26}-\epsilon g_{15} (2v_5-v_1)+\epsilon g'_{15}=0,\nonumber\\
(A^g)_{27,16}&=-g_{13}-g_{14} -g_{16}g_{26}-\epsilon g_{16}(v_1-v_2+v_3-v_4+v_5)
+\epsilon g'_{16}=0,\nonumber\\
(A^g)_{27,17}&=-g_{14}-g_{15}-g_{17}g_{26}-\epsilon g_{17}(-v_1+v_4)+\epsilon g'_{17}=0,
\nonumber\\
(A^g)_{27,18}&=-g_{16}-g_{18}g_{26}-\epsilon g_{18}(v_5+v_1-v_3+v_6)+\epsilon g'_{18}=0,\nonumber\\
(A^g)_{27,19}&=-g_{16}-g_{17}-g_{19}g_{26}-\epsilon g_{19}(-v_1+v_3-v_4+v_5)
+\epsilon g'_{19}=0,\nonumber\\
(A^g)_{27,20}&=-g_{18}-g_{20}g_{26}-\epsilon g_{20}(v_1+v_5-v_6)+\epsilon g'_{20}=0,\nonumber\\
(A^g)_{27,21}&=-g_{18}-g_{19}-g_{21}g_{26}-\epsilon g_{21}(v_5-v_1+v_2-v_3+v_6)
+\epsilon g'_{21}=0,\nonumber\\
(A^g)_{27,22}&=-g_{20}-g_{21}-g_{22}g_{26}-\epsilon g_{22} (v_5-v_1+v_2-v_6)
+\epsilon g'_{22}=0,\nonumber\\
(A^g)_{27,23}&=-g_{21}-g_{23}g_{26}-\epsilon g_{23}(v_5-v_2+v_6)+\epsilon g'_{23}=0,
\nonumber\\
(A^g)_{27,24}&=-g_{22}-g_{23}-g_{24}g_{26}-\epsilon g_{24}(v_5-v_2+v_3-v_6)
+\epsilon g'_{24}=0,\nonumber\\
(A^g)_{27,25}&=-g_{24}-g_{25}g_{26}-\epsilon g_{25}(-v_3+v_4+v_5)+\epsilon g'_{25}=0,\nonumber\\
(A^g)_{27,26}&=-g_{25}-g_{26}^2-\epsilon g_{26}(2v_5-v_4)+\epsilon g'_{26}=0.
\end{align}
Here, we denote $g_{27,i}$ as $g_{i}$.

The zeroth order solutions are expressed in terms of $s_{26}^{(0)}$ as follows:
\begin{align}
    s_{1}^{(0)}&={1\over 117}(-1+252t) (s_{26}^{(0)})^{16},\quad
    s_{2}^{(0)}={1\over 78}(1-213t) (s_{26}^{(0)})^{15},\nonumber \\
    s_{3}^{(0)}&={1\over 78}(-1+135t) (s_{26}^{(0)})^{14},\quad
    s_{4}^{(0)}={1\over 78}(1-57t) (s_{26}^{(0)})^{13},\nonumber\\
    s_{5}^{(0)}&={-1\over 78}(1+21t) (s_{26}^{(0)})^{12},\quad
    s_{6}^{(0)}={1\over 78}(-1+57t) (s_{26}^{(0)})^{12},\nonumber\\
    s_{7}^{(0)}&={1\over 78}(1+99t) (s_{26}^{(0)})^{11},\quad
    s_{8}^{(0)}={1\over 39}(1-18t) (s_{26}^{(0)})^{11},\nonumber\\
    s_{9}^{(0)}&=-{1\over 26}(1+21t) (s_{26}^{(0)})^{10},\quad
    s_{10}^{(0)}={1\over 39}(-1+18t) (s_{26}^{(0)})^{10},\nonumber\\
    s_{11}^{(0)}&={1\over 78}(5+27t) (s_{26}^{(0)})^{9},\quad
    s_{12}^{(0)}={1\over 39}(1-18 t) (s_{26}^{(0)})^{9},\nonumber\\
    s_{13}^{(0)}&=-{1\over 78}(5+27t) (s_{26}^{(0)})^{8},\quad
    s_{14}^{(0)}={1\over 78}(-7+9t) (s_{26}^{(0)})^{8},\nonumber\\
    s_{15}^{(0)}&={1\over 39}(-1+18t) (s_{26}^{(0)})^{8},\quad
    s_{16}^{(0)}={1\over 13}(2+3t) (s_{26}^{(0)})^{7},\nonumber\\
    s_{17}^{(0)}&=-{3\over 26}(-1+5t) (s_{26}^{(0)})^{7},\quad
    s_{18}^{(0)}=-{1\over 13}(2+3t) (s_{26}^{(0)})^{6},\nonumber\\
    s_{19}^{(0)}&={1\over 26}(-7+9t) (s_{26}^{(0)})^{6},\quad
    s_{20}^{(0)}={1\over 13}(2+3t) (s_{26}^{(0)})^{5},\nonumber\\
    s_{21}^{(0)}&={1\over 26}(11-3t) (s_{26}^{(0)})^{5},\quad
    s_{22}^{(0)}=-{3\over 26}(5+t) (s_{26}^{(0)})^{4},\nonumber\\
    s_{23}^{(0)}&={1\over 26}(-11+3t) (s_{26}^{(0)})^{4},\quad
    s_{24}^{(0)}= (s_{26}^{(0)})^{3},\nonumber\\
    s_{25}^{(0)}&=- (s_{26}^{(0)})^{2},\quad
    p(z)=t (s_{26}^{(0)}(z))^{12},
    \label{eq:zeroth1}
\end{align}
where $t$ is 
\begin{align}
t=\frac{1}{9}\qty(-45\pm26\sqrt{3}).
\end{align}

\section{Integral formulae for the calculation of $Q_6$}\label{q6formula}

We calculate the sixth period (\ref{Q6inC6}) by using the following formulae. These are derived by substituting the explicit form of $s_{26}^{(0)}(z)$ in (\ref{s260}) into the integrals, writing them as linear combinations of $J(-\frac{5}{h}+m,-6+n(hM))$, and using the recurrence relation in (\ref{A3}) so that the expressions are factorized by $J(-\frac{5}{h},-6)$. Note that the overall factor $t^{5/h}$ appears in every formula.\\
{\tiny
\begin{align}
  \oint_C\dd z\frac{(s_{26}^{(0)''})^3}{(s_{26}^{(0)})^8}
  &=
  -\frac{2 t^{5/h}J(-\frac{5}{h},-6) (h M-5) (h M-1) (2 h M-5)}{(h+1) (h+5) (2 h+5) (3 h+5) (4 h+5)} \left(h \left(h \left((h (h+12)+59) M^2+(h (h+13)-3) M+h-14\right)-5 (31 M+3)\right)+100\right),
  \nonumber\\
  \oint_C\dd z\frac{(s_{26}^{(0)''})^2}{z^2(s_{26}^{(0)})^7}
  &=
  -\frac{t^{5/h}J(-\frac{5}{h},-6) (h M-5)}{(h+5) (2 h+5) (3 h+5)} (h (h M (5 h (M+1)+51 M+28)+h-5 (29 M+9))+100),
  \nonumber\\
  \oint_C\dd z\frac{s_{26}^{(0)''}(s_{26}^{(0)})^{(4)}}{(s_{26}^{(0)})^7}
  &=
  \frac{t^{5/h}J(-\frac{5}{h},-6) (h M-5) (h M-1)}{(h+1) (h+5) (2 h+5) (3 h+5) (4 h+5)} \left[h \left(h \left(4 h^4 M^2 (M+1)+3 h^3 M (M+1) (9 M+8)+h^2 (M+1) (M (68 M-3)-76)
  \right.\right.\right.
  \nonumber\\
  &\left.\left.\left.\quad
  -3 h (M (M (369
   M+49)+106)+42)+6075 M^2-65 M+100\right)+150 (1-71 M)\right)+6000\right],
  \nonumber\\
  \oint_C\dd z\frac{s_{26}^{(0)''}}{z^4(s_{26}^{(0)})^6}
  &=
  -\frac{2 t^{5/h}J(-\frac{5}{h},-6) (h (3 M+2)-5) }{h+5},
  \nonumber\\
  \oint_C\dd z\frac{((s_{26}^{(0)})^{(3)})^2}{(s_{26}^{(0)})^7}
  &=
  -\frac{ t^{5/h}J(-\frac{5}{h},-6) (h M-5) (h M-1)}{(h+1) (h+5) (2 h+5) (3 h+5) (4 h+5)}\left[h \left(h \left(4 h^4 M^2 (M+1)+h^3 M (M+1) (13 M+24)+2 h^2 (M+1) (M (34 M+9)-38)
  \right.\right.\right.
  \nonumber\\
  &\left.\left.\left.\quad
  +h (M (M (923
   M-217)-367)-91)+10 (4-495 M) M+310\right)+8425 M+325\right)-4500\right],
  \nonumber\\
  \oint_C\dd z\frac{(s_{26}^{(0)})^{(3)}}{z^3(s_{26}^{(0)})^6}
  &=
 -\frac{6 t^{5/h}J(-\frac{5}{h},-6) \left(h^2 (M (7 M+9)+4)-15 h (2 M+1)+25\right) }{(h+5) (2 h+5)},
  \nonumber\\
  \oint_C\dd z\frac{(s_{26}^{(0)})^{(4)}}{z^2(s_{26}^{(0)})^6}
  &=
  \frac{6t^{5/h}J(-\frac{5}{h},-6)}{(h+5) (2 h+5) (3
   h+5)} \left(5 h^4 M^2 (M+1)-h^3 (M (M (61 M+109)+96)+24)+5 h^2 (M (88 M+71)+27)-25 h (35 M+11)+500\right),
  \nonumber\\
  \oint_C\dd z\frac{(s_{26}^{(0)})^{(6)}}{(s_{26}^{(0)})^6}
  &=
  -\frac{6t^{5/h}J(-\frac{5}{h},-6) (h M-5) (h M-1)}{(h+1) (h+5) (2 h+5) (3 h+5) (4 h+5)}\left[4 h^6 M^2 (M+1)+h^5 M (M+1) (55 M+24)-h^4 (M+1) (5 M (188 M+9)+76)
  \right.
  \nonumber\\
  &\left.\quad
  +h^3 (M (M (1889 M+3185)+3980)-196)-5 h^2
   (M (2271 M+55)+904)+200 h (113 M-22)-15000\right],
    \nonumber\\
  \oint_C\dd z\frac{1}{z^6(s_{26}^{(0)})^{5}}&=
  t^{5/h}J(-\frac{5}{h},-6).
\end{align}
}
We used these formulae with $h=12$.

\bibliographystyle{utphys}
\bibliography{refs.bib}

\end{document}